\documentstyle[twoside,fleqn,espcrc2]{article}

\begin{document}

\title{Hadronic Spectroscopy from the Lattice: Glueballs and Hybrid Mesons}

\author{Chris Michael\address{
Division of Theoretical Physics,
 Department of Mathematical Sciences, \\
University of Liverpool, Liverpool L69 3BX, UK}%
  }

\begin{abstract}

 Lattice QCD determinations appropriate to hadron spectroscopy are
reviewed with emphasis on the  glueball and hybrid meson states in the
quenched approximation. Hybrids are  discussed for heavy and for light
quarks. The effects of sea quarks (unquenching) are  explored.

\end{abstract}

\maketitle

\section{INTRODUCTION}

   Quantum Chromodynamics is generally acknowledged to be the theory of
hadronic  interactions. Its perturbative features are well understood
and provided the  main motivation for its adoption. To give a complete
description of hadronic  physics, it is essential to develop
non-perturbative methods to handle QCD. QCD is a quantum field theory
which  needs to be regulated in order to have a well defined
mathematical approach.  Inevitably such a regularisation will destroy
some of the symmetries of QCD.  For example  dimensional regularisation
is often used in perturbative  studies and it breaks the 4-dimensional
Lorentz invariance. Similarly lattice regularisation as proposed by
Wilson~\cite{wilson} breaks Lorentz invariance since a hypercubic
lattice of space-time points is invoked. The key feature of Wilson's
proposal  is that gauge invariance is exactly retained. Then the
approach to the continuum limit (as the lattice spacing $a$ is reduced
to zero) can be shown to be  well defined. Using Monte Carlo methods to 
explore lattice QCD, reliable predictions can now be made for continuum 
quantities as I shall discuss. Moreover, in the continuum limit, the Lorentz 
invariance is found to be fully restored.

  Lattice QCD needs as input the quark masses and an overall scale
(conventionally  given by $\Lambda_{QCD}$). Then any Green function can
be evaluated by taking an average of suitable combinations of the
lattice fields in the vacuum samples. This allows masses to be studied 
easily and matrix elements (particularly those of weak or
electromagnetic currents)  can be extracted straightforwardly.
  Unlike experiment, lattice QCD can vary the quark masses and can also 
explore different boundary conditions and sources. This allows a wide
range of  studies which can be used to diagnose the health of
phenomenological models as well as casting light on experimental data.

 It is worth recalling the weakness as well as the strength of the
lattice approach  to QCD. Any quantity which can be expressed as a
vacuum expectation  value of fields can be extracted straightforwardly
in lattice studies. Thus  masses and matrix elements of operators can be
determined. What is not  so easy is to explore hadronic decays. This is
difficult because the  lattice, using Euclidean time, has no concept of
in and out states. About the  only feasible strategies are to evaluate the
mixing between states of the same  energy - so giving some information
on on-shell hadronic decay amplitudes, or to make a model 
dependent analysis~\cite{cmdecay} of lattice results.

 One very special case is of considerable interest: this is {\em
quenched} QCD where the sea-quark masses are taken as infinite. This
suppresses  quark loops in the vacuum completely, leaving just the full
non-perturbative  gluonic interactions. This gluonic vacuum turns out to
reproduce most of  the salient features of QCD. It is also a very
convenient approximation to  use for comparison with phenomenological
models. Quenched QCD is computationally rather easy to study and the
precise results allow the continuum limit to be extracted  reliably. 
Studies with sea quark effects included (known as dynamical fermion 
studies) are computationally much more demanding. I discuss the current
situation  in this area in the last section.

\section{GLUEBALLS}

Glueballs are defined to be hadronic states made primarily from gluons.
The full non-perturbative gluonic interaction is included in quenched
QCD.  In the quenched approximation, there is no mixing between such
glueballs  and quark - antiquark mesons. A study of the glueball
spectrum in quenched QCD  is thus of great value. This will allow
experimental searches to be  guided as well as providing calibration for
models of glueballs.

In principle, lattice QCD can study the meson spectrum as the sea quark
mass  is decreased towards experimental values. This will allow the
unambiguous glueball states in the quenched approximation to be tracked
as the sea quark effects are increased.  It may indeed turn out that no
meson in the physical spectrum is primarily a glueball - all states are 
mixtures of glue,  $q \bar{q}$, $q \bar{q} q \bar{q}$, etc.  Studies 
conducted so far show no significant change of the glueball spectrum as
dynamical quark effects are added - but  the sea quark masses used are
still rather large~\cite{sesam}.

In lattice studies, dimensionless ratios of  quantities are obtained. To
explore the glueball masses, it is appropriate to combine  them with
another very accurately measured quantity to have a dimensionless 
observable. Since the potential between static quarks is very accurately
measured from the lattice (see the next section for more details), it is
 now conventional to use $r_0$ for this comparison.  Here $r_0$ is
implicitly defined by $r^2 dV(r)/dr = 1.65$ at $r=r_0$ where $V(r)$ is 
the potential energy between static quarks which is easy to determine
accurately  on the lattice. In practice $r_0$ may be related to the
string tension  $\sigma$ by $r_0 \sqrt{\sigma}=1.18$. Taking the
conventional value of $\sqrt{\sigma}=0.44$ GeV then yields $r_0=2.68\
{\rm GeV}^{-1}$=0.53 fm. Taking the potential $V(r)$  from $b\bar{b}$
spectra suggests that $r_0 \approx 0.5$ fm which is  a very similar
estimate.

 Theoretical analysis  indicates that for the quenched approximation 
the dimensionless ratio $mr_0$ will differ from the continuum  limit
value by corrections of order $a^2$.  Thus in Fig.~1 the masses are
plotted versus the lattice spacing $a^2$ for the $J^{PC}$=$0^{++}$ and
$2^{++}$ glueballs. The straight lines then show the continuum limit
obtained  by extrapolating to $a=0$. As can be seen, there is
essentially no need for data  at even smaller $a$-values to further fix
the continuum value. The values shown  correspond to
$m(0^{++})r_0=4.33(5)$ and $m(2^{++})r_0=6.0(6)$.  Since several
lattice groups~\cite{DForc,MTgl,ukqcd,gf11} have measured these 
quantities, it is reassuring to see that the purely lattice observables
are in  excellent agreement. The publicised difference of quoted
$m(0^{++})$ from  UKQCD~\cite{ukqcd} and GF11~\cite{gf11} comes entirely
from relating quenched lattice  measurements to values in GeV as I now
discuss.

\begin{figure}[bt] 
\vspace{8cm} 
\includegraphics{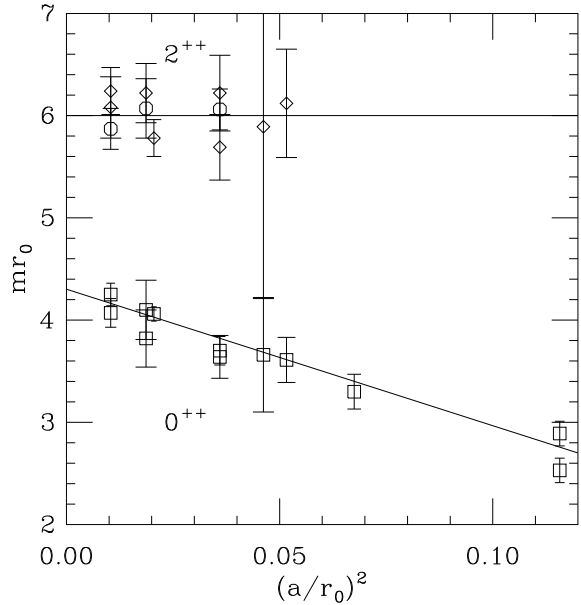}
 \caption{ The value of mass of the  $J^{PC}=0^{++}$ and $2^{++}$
glueball states from refs{\protect\cite{DForc,MTgl,ukqcd,gf11}} in units
of $r_0$ where $r_0 \approx 0.5$ fm. The $T_2$ and $E$ representations
are shown  by  octagons  and    diamonds respectively and their agreement
indicates the restoration of rotational invariance for the $2^{++}$
state.  The straight lines  show  fits describing the  approach to the
continuum limit as $a \to 0$.
   }
\end{figure}

In the quenched approximation, different hadronic observables differ
from experiment  by factors of up to 10\%. Thus using one quantity or
another to set the scale, gives an overall systematic error.  Here I
choose to set the scale by taking the conventional value of the string
tension, $\sqrt{\sigma}=0.44$ GeV, which then corresponds to
$r_0^{-1}=373$ MeV. An overall systematic error of 10\% is then to be
included to any  extracted mass. This yields $m(0^{++})=1611(30)(160)$
MeV and  $m(2^{++})=2232(220)(220)$ MeV where the second error is the
systematic  scale error. Note that these are glueball masses in the
quenched approximation -  in the real world significant mixing with $q
\bar{q}$ states could modify these values substantially.

Recently a lattice approach using a large spatial lattice spacing with
an  improved action and a small time spacing has been used to study
glueball  masses. The results~\cite{mpglue} are that
$m(0^{++})r_0=3.98(15)$,  $m(2^{++})r_0=5.85(2)$, $m(1^{+-})r_0=7.21(2)$
and $m'(2^{++})r_0=8.11(4)$. There remains a small discrepancy with the
result for the $0^{++}$ glueball obtained above ($m(0^{++})r_0=4.33(5)$)
from lattice spacings much closer to the continuum limit. This has
recently been explored  in more detail~\cite{mpglue2} and modifications
proposed to the technique  which allow better control of the error in
extrapolating to  the continuum limit. 

I have focussed on the scalar and tensor glueball results because these
are the  lightest and best measured states in lattice studies. The
glueball spectrum has been  extracted for all $J^{PC}$
values~\cite{MTgl,ukqcd}. Some of these results are shown in Fig.~2. 
One signal of great interest would be  a glueball with $J^{PC}$ not
allowed for $q \bar{q}$ - a spin-exotic glueball or {\em oddball}. These
states are shown  in Fig.~2 to be high lying: at least above
$2m(0^{++})$. Thus they are  likely to be in a region very difficult to
access unambiguously by experiment.

\begin{figure*}[t]
\vspace{10.3cm} 
\includegraphics{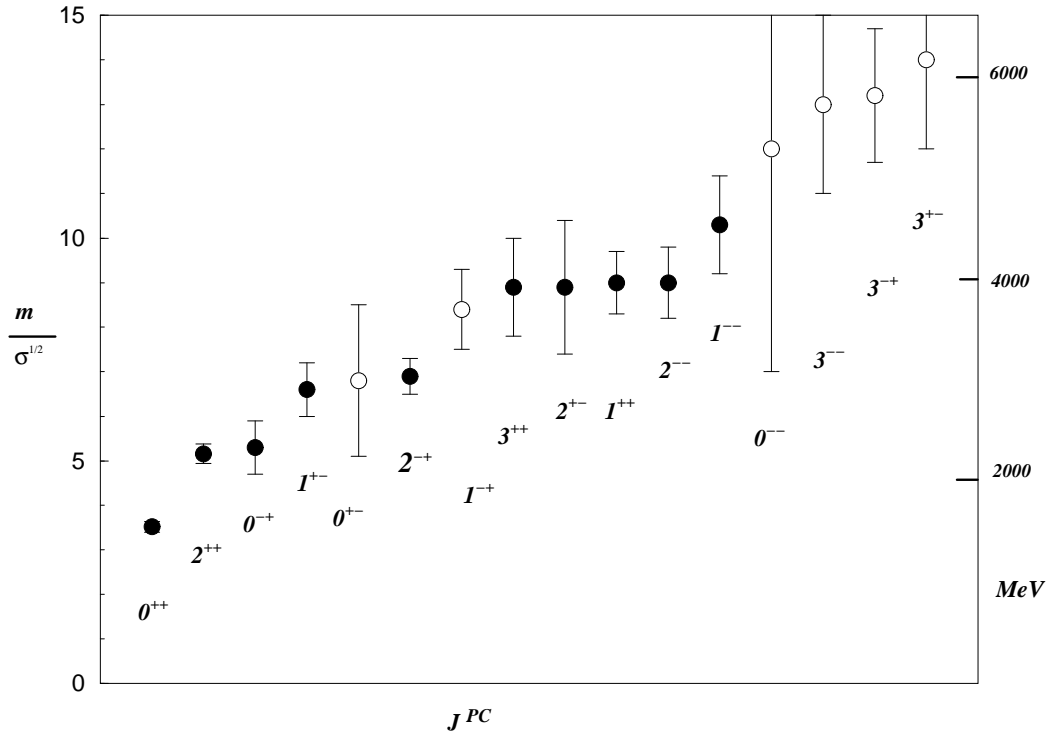}
 \caption{ The  mass of the  glueball states  with quantum  numbers
$J^{PC}$ from ref{\protect\cite{ukqcd}}.  The scale is set by
${\protect\sqrt{\sigma}} \approx 440 $ MeV which yields the right hand
scale in MeV. The solid circles represent mass determinations whereas
the open circles  are upper limits. Note that excited states in several
cases  have been determined, such as $  m'(0^{++})/\sqrt{\sigma} =6.3(4)$
from ref{\protect\cite{MTgl}}. 
   }
\end{figure*}

 The only other candidate for a relatively light glueball is the
pseudoscalar. Values quoted of $m(0^{-+})r_0= 5.6(6),\ 7.1(1.1)$ and
5.3(6) from refs\cite{MTgl,ukqcd} suggest  an average of 6.0(1.0), close
to  the tensor glueball mass. This is confirmed by a study using
asymmetric lattices~\cite{mpglue2} which obtains a pseudoscalar glueball
mass ratio to the tensor glueball of $m(0^{-+})/m(2^{++})=1.17(9)$.

 Within the quenched approximation, the glueball states are unmixed 
with $q \bar{q},\ q \bar{q} q \bar{q}$, etc. Furthermore, the  $q
\bar{q}$ states are `magically mixed' in the quenched approximation. 
Once quark loops are allowed in the vacuum, for the favour-singlet
states of any given $J^{PC}$,  there will be mixing between the  $s
\bar{s}$ state, the  $u \bar{u}+d \bar{d}$ state  and the glueball. For
the $0^{++}$ case, as a guide to this mixing,  the quenched  mass
spectrum of $q \bar{q}$ states has been determined on a lattice and  the
scalar mesons  are found to lie somewhat lighter than the tensor
states~\cite{hybrid}. These  $2^{++}$ mesons are experimentally
approximately `magically mixed' so  will be quite close to the quenched
mass determination. This suggests  that the quenched scalar masses from
the lattice are at around 1.2 GeV and 1.5 GeV. An independent
study~\cite{weinss,weinssg}  suggests that the scalar $s \bar{s}$ state
is about 200 MeV lighter than the glueball which is a broadly compatible
conclusion.  Thus the glueball, at around 1.6 GeV, lies heavier than the
$q\bar{q}$ scalar states.

A first attempt~\cite{weinssg} has been made to measure, on a quenched
lattice, the  mixing matrix element between the glueball and the scalar
$q\bar{q}$ states.  The mixing matrix element between the $s\bar{s}$
meson and the glueball  is extracted as $E(s)=43(31)$ MeV. The mixing
matrix elements with the glueball are found to be similar for light and
$s\bar{s}$ quarks (actually $E(u,d)/E(s) \approx 1.2$).  With this
lattice input, a mixing model~\cite{weinmix,weinssg} allows the content
of the experimental scalar states at 1390, 1500 and 1710 MeV to be
assigned. Other,  more phenomenological mixing models have also been
proposed~\cite{close}. In these mixing scenarios none  of the
experimental states  is predominantly an unmixed  glueball.

As well as this mixing of the glueball with $q \bar{q}$ states, there
will be  mixing  with $q \bar{q} q \bar{q}$ states which will be
responsible for the  hadronic decays. Because of the limitations of
lattice techniques, it is  only possible to estimate these decay matrix
elements when the initial and final  states have the same energy and
this has been applied  to the  decay matrix element between a $0^{++}$
glueball and two pseudoscalar $q\bar{q}$  mesons. This then allows an
estimate of the width which would arise  when going beyond the quenched
approximation.  A first attempt to study this~\cite{gdecay} yields
estimated widths of order 100 MeV. Even though this lattice calculation
is very exploratory, it does indicate that very wide widths to two
pseudoscalars are not expected. A more realistic study  would involve
taking account of mixing with the $n \bar{n}$ and $s \bar{s}$ scalar
mesons as  well as the  decay channels.

\section{HEAVY QUARK INTERACTIONS}

In the limit  $m_Q \to \infty$, the heavy quark effective theory
describes a universal behaviour. For finite $m_Q$, corrections of order
$1/m_Q$  are expected. The simplest way to study the heavy quark limit
on a lattice is to  use static quarks. The potential energy $V(R)$
between a static quark and antiquark  at separation $R$ is readily
obtained. Then for heavy quarks, one may solve for  the spectra in this
potential using the Schr\"odinger equation in the  adiabatic
approximation.  This is expected to be a good approximation for
determining the  spectrum of $b \bar{b}$ mesons. The quenched lattice
potential is well measured and is found to have a form parametrised by 
 \begin{equation}
 V(R) = V_0 - {e \over R} + \sigma R
 \end{equation}  
 where $e$ is the coefficient of the Coulomb term and $\sigma$ is the
string tension. This expression shows that the potential  continues to
increase as $R$ is increased - this is confinement.

 A comparison from ref\cite{pm} of the spectrum in the quenched lattice
potential with  the $\Upsilon$ states is shown in  Fig.~3. The lattice
result is qualitatively similar to the experimental $\Upsilon$ spectrum.
The main difference is that the Coulombic part ($e$) is effectively too
small (0.28 rather than 0.48). This produces~\cite{pm} a ratio of   mass
differences $(1P-1S)/(2S-1S)$ of 0.71 to be compared  with the
experimental ratio of 0.78.  This difference in $e$ is understandable as
a consequence of the one gluon exchange (Coulombic) force at short
distances which would be increased by $33/(33-2N_f)$ in perturbation
theory in full QCD compared to quenched QCD. We will return to discuss
this.

\begin{figure}[bt!] 
\vspace{8cm} 
\includegraphics{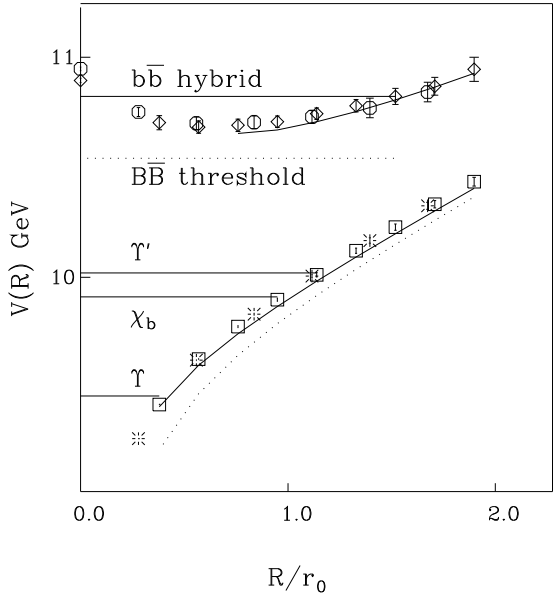}
 \caption{ Potentials $V(R)$ in GeV between static quarks  at separation
$R$ (in units of $r_0 \approx 0.5$ fm) for the  ground state (square and
{\tt *}) and for the $E_u$ symmetry which corresponds to the first
excited state of the gluonic flux (octagons and diamonds). Results  from
 the quenched calculations of ref{\protect\cite{pm}} are shown by
symbols corresponding to different lattice spacings. For the ground
state  potential the continuous curve is an interpolation of the lattice
data while the dotted  curve with enhanced Coulomb term fits the
spectrum and yields the masses shown. The lightest hybrid level in the
excited gluonic potential is  also shown.
   }
\end{figure}

The situation of a static quark and antiquark is a very clear case in
which to discuss hybrid  mesons which  have excited gluonic
contributions. A discussion of the colour  representation of the quark
and antiquark is not useful since they are at different space positions
and the combined colour is not gauge invariant. A better criterion is 
to focus on the spatial symmetry of the gluonic flux. As well as the
symmetric ground state of the colour flux between two  static quarks,
there will be excited states with different symmetries.  These were
studied on a lattice~\cite{liv} and the conclusion was that the $E_u$
symmetry (corresponding to flux  states from an  operator which is the
difference of U-shaped paths from quark to antiquark of the form $\,
\sqcap - \sqcup$) was the lowest lying gluonic excitation.  Results  for
this potential are shown in Fig.~3.

This gluonic excitation in the $E_u$ representation (or $\Pi_u$ in the
continuum) corresponds to a component of angular momentum of one unit
along the quark antiquark axis. Then one can solve for the spectrum of
hybrid mesons using the Schr\"odinger equation in the adiabatic
approximation. The spatial wave function necessarily has non zero
angular momentum and  the lightest states correspond to $L^{PC}=1^{+-}$
and $1^{-+}$. Combining with the  quark and antiquark spins then
yields~\cite{liv} a set of 8 degenerate hybrid states with
$J^{PC}=1^{--},\ 0^{-+},\ 1^{-+},\ 2^{-+}$ and    $1^{++},\ 0^{+-},\
1^{+-},\ 2^{+-}$  respectively. These contain the  spin-exotic states
with $J^{PC}=  1^{-+},\ 0^{+-}$ and $2^{+-}$ which will be of special
interest. 

 Since the lattice calculation of the ground state and hybrid masses
allows  a direct prediction for their difference, the result for this
8-fold degenerate hybrid level is illustrated in Fig.~3  and
corresponds~\cite{pm} to masses of 10.81(25) GeV for $b\bar{b}$ and 
4.19(15) GeV for $c \bar{c}$ when using the Schr\"odinger equation in the
Born-Oppenheimer  approximation to determine the spectrum. Here the
errors take into account the uncertainty in setting the ground state
mass using the quenched potential as discussed above. Recently a
different lattice technique~\cite{morn} has been used to  explore in
detail the excited gluonic levels in the quenched approximation. The
results above are confirmed and the value quoted for the lightest
hybrid meson is $(m_{\rm Hybrid}-m_{\Upsilon})r_0=3.166(3)$ for
$b \bar{b}$ where the error  does not take into account unquenching

 The quenched lattice results, after adjusting to take account  of the
measured $\bar{b} b $ spectrum,  suggest that the lightest hybrid mesons
lie  above the open $B \bar{B}$ threshold by about 270 MeV.  This can
also be studied by  comparing directly the lattice hybrid masses with
twice the quenched lattice  masses for the $B$ meson~\cite{sommer}.
Using quenched  results from the smallest lattice spacing ($\beta=6.2$)
available with clover-improved fermions~\cite{ewing} yields $m_{\rm
Hybrid} - 2 m_{B} \approx  140(80)$ MeV. This estimate is somewhat
smaller than that  obtained above.  In both cases, however, the hybrid 
levels lie above the open threshold  and are likely to be relatively
wide  resonances. Another consequence~\cite{pm} is that the very flat
potential implies a very extended wavefunction. This is illustrated, for
 example, in ref~\cite{morn} which shows that the 
hybrid wavefunction  remains significant  out to  a radius of 1 fm.
This has the implication that the
wavefunction at the origin will be small,  so hybrid vector states will
be weakly produced from $ e^+ e^-$.

 It would be useful to explore the splitting among the 8 degenerate
$J^{PC}$ values obtained. This could come from different excitation 
energies in the $L^{PC}=1^{+-}$ (magnetic) and $1^{-+}$
(pseudo-electric) gluonic excitations, spin-orbit terms, as well as
mixing between hybrid states and $Q\bar{Q}$ mesons with non-exotic spin.
One way to study this on a lattice is to use the  NRQCD formulation
which describes non-static heavy quarks which propagate 
non-relativistically. Preliminary results for hybrid excitations from
several  groups~\cite{manke,collins,morn} give at present similar
results to those with the Born-Oppenheimer approximation in the static
approximation as described above. There is some evidence that  the
magnetic excitations (which include the $1^{-+}$ spin  exotic) are
lighter than the pseudo-electric ones (which include the  $0^{+-}$ spin
exotic).  For example ref\cite{morn} obtains
 $(m(1^{+-})-m_{\Upsilon})r_0=3.29(5)$  and
 $(m(0^{-+})-m_{\Upsilon})r_0=3.51(8)$, to be compared with the static
approximation,  discussed above, which has degenerate levels with  a
value of 3.166(3). It has to be remembered that these are all quenched
determinations,  so there are systematic errors of order 10\% from
unquenching. This implies  that the NRQCD estimate for the lightest
spin-exotic hybrid would be the  $1^{+-}$ state at a mass of
10.69(13) GeV.


\section{LIGHT QUARK INTERACTIONS }

 Unlike very heavy quarks, light quark propagation in the gluonic vacuum
sample is very computationally intensive - involving inversion of huge
($10^9 \times 10^9$) sparse matrices. Current computer power is 
sufficient to study light quark physics thoroughly in the quenched 
approximation. The state of the art~\cite{yoshie} is the Japanese CP-PACS
Collaboration  who are able to study a range of large lattices (up to
about $64^4$) with a range of light quark masses. Qualitatively the 
meson and baryon spectrum of states made of  light and strange quarks is 
reproduced with discrepancies of order 10\% in the quenched approximation.

\begin{figure}[bt!] 
\vspace{8cm} 
\includegraphics{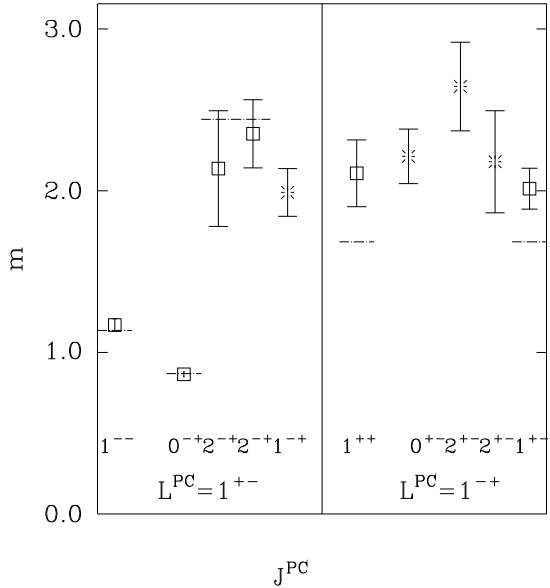}
 \caption{ The masses in  GeV of states of $J^{PC}$ built from hybrid
operators with strange quarks, spin-exotic ({\tt *}) and non-exotic
(squares). The dot-dashed lines are the mass values found for 
$s\bar{s}$ operators. Results from 
 ref{\protect\cite{hybrid}}. 
   }
\end{figure}

 Here I will focus on hybrid mesons made from light quarks.  There will
be no mixing with $q \bar{q}$ mesons for  spin-exotic hybrid mesons  and
these are of special interest. The first study of this area was by the 
UKQCD Collaboration~\cite{hybrid} who used operators motivated by the 
heavy quark studies referred to above. Using non-local operators, they
studied  all 8 $J^{PC}$ values coming from $L^{PC}=1^{+-}$ and $1^{-+}$
excitations. The  resulting mass spectrum is shown in Fig.~4 where the
$J^{PC}=1^{-+}$ state  is seen to be the lightest spin-exotic state with
a statistical significance of 1 standard deviation. The statistical
error on the mass of this lightest spin-exotic meson  is 7\% but, to
take account of systematic errors from the lattice determination, a 
mass of 2000(200) MeV is quoted for this hybrid meson with $s \bar{s}$
light quarks. Although not  directly measured, the corresponding light
quark hybrid meson would be expected to be around 120 MeV lighter. In
view of the   small statistical error, it seems unlikely that the
$1^{-+}$ meson in the quenched approximation could lie as light as 1.4
GeV where there are experimental  indications for such a
state~\cite{bnl}. Recently experimental evidence has also been
presented~\cite{hyb16}  for another  $1^{-+}$ meson at 1.6 GeV.

One feature clearly seen in Fig.~4 is that non spin-exotic mesons
created  by hybrid meson operators have  masses  which are very similar
to those found when the states are created by $q \bar{q}$ operators.
This suggests that there is  quite strong coupling between hybrid and $q
\bar{q}$ mesons even in the quenched approximation. This would imply that 
the $\pi(1800)$ is unlikely to be a pure hybrid, for example.

A second lattice group has also evaluated hybrid meson spectra from
light quarks from quenched lattices. They obtain~\cite{milc} masses of
the $1^{-+}$ state with statistical and various systematic errors of 
1970(90)(300) MeV, 2170(80)(100)(100) MeV and 4390(80)(200) MeV for $n
\bar{n}$,  $s \bar{s}$ and $c \bar{c}$ quarks respectively. For the 
$0^{+-}$ spin-exotic state they have a noisier signal but evidence that
it is heavier. They also explore mixing matrix elements between
spin-exotic hybrid  states and 4 quark operators.

\begin{figure}[bt!] 
\vspace{8cm} 
\includegraphics{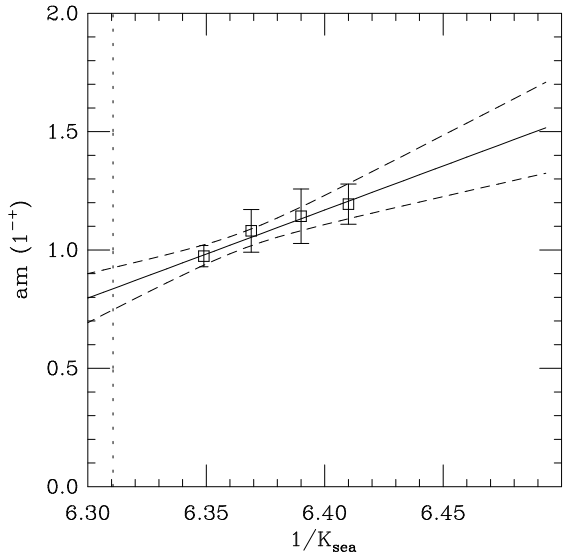}
 \caption{ The extrapolation in sea-quark mass (in lattice units) from 
 ref{\protect\cite{sesamhyb}} for the $1^{-+}$ hybrid meson. The dotted
vertical  line corresponds to light sea quarks.
   }
\label{fig.sesam}
\end{figure}

 Recently a first attempt has been made~\cite{sesamhyb} to determine the
hybrid meson spectrum using  full QCD. The sea-quarks used have several
different masses and an extrapolation  is made to the limit of physical
sea-quark masses, yielding a mass of 1.9(2) GeV for the lightest 
spin-exotic hybrid meson, which they again find to be the $1^{-+}$. In
principle this  calculation should take account of sea-quark effects
such as the mixing  between such a hybrid meson and $q \bar{q} q
\bar{q}$ states such as $\eta \pi$. As illustrated in
Fig.~\ref{fig.sesam}, the calculations are performed for  quite heavy
sea quarks (the lightest being approximately the  strange quark mass)
and then a linear extrapolation is made. It is  quite possible, however,
that such mixing effects turn on non-linearly as the sea-quark  masses
are reduced. The systematic error from this possibility is  difficult to
quantify.  

The three independent lattice calculations of the light hybrid spectrum
are  in good agreement with each other. They imply that the natural
energy  range for spin-exotic hybrid mesons is around 1.9 GeV. The
$J^{PC}=1^{-+}$  state is found to be lightest. It is not easy to
reconcile these lattice results  with experimental indications for
resonances at 1.4 GeV and 1.6 GeV. Mixing  with  $q \bar{q} q \bar{q}$
states such as $\eta \pi$ is not included for realistic quark masses in
the  lattice calculations. This can be interpreted, dependent on one's
viewpoint,  as either that the lattice calculations  are incomplete or
as an indication that the experimental states may have an  important
meson-meson component in them.

\section{TOWARDS FULL QCD}

 Algorithms exist which allow lattice simulation of full QCD with  sea
quarks of mass $m_{sea}$. This study needs lots of CPU power  since the
sea quark loops in the vacuum are represented effectively as  a long
range interaction between the gluonic degrees of freedom. Most  studies
to date have been exploratory with sea quark masses above the  strange
quark mass. In this regime, very little change from the quenched
approximation is seen in  physical predictions from the lattice. 

One area where specific changes are  expected is for the potential
between heavy quarks. At small separation $R$,  the Coulombic term is
expected to increase in strength and indeed evidence of this  has been
reported~\cite{sesamv,ukqcddf,cppacs}. At larger $R$,  signs of string
breaking are expected since a light quark antiquark pair can be produced
 from the vacuum to yield two $Q \bar{q}$ mesons with energy independent
of $R$  at large separation. This has been explored by the SESAM,  UKQCD
and CP-PACS collaborations  and little sign of the effect is
seen~\cite{ukqcddf,sesamv,cppacs}. 

It has been very difficult to see unambiguous signs of sea  quark
effects in the hadron spectrum. The CP-PACS collaboration~\cite{cppacsh}
have recently reported that they see evidence that, for two flavours of
light  quark, results are closer to experiment than from quenched
studies. Even their study, however, is restricted to sea-quark masses
which are not appreciably lighter  than the strange quark mass.  Now, 
it is possible that such sea-quark effects turn on non-linearly  as the
sea quark mass is reduced. As an example, in current studies the $\pi +
\pi$ P-wave is heavier than the $\rho$ meson so the $\rho$  cannot 
decay. Another example is that glueball mixing with $q\bar{q}$  scalar
mesons may depend quite sensitively on their masses, so as the sea 
quark mass is reduced, the scalar meson mass will move relative  to the
glueball   mass so causing a non-linear change in the mixing strength.

Further work is needed to reduce the sea quark mass and to increase  the
lattice size. Dedicated computing  power of several hundreds of Gflops
is available to lattice collaborations  and, with future plans for
Tflops  machines, progress in this area should now be possible.

\section{SUMMARY}

 Quenched lattice QCD is well understood and accurate predictions in the
continuum limit  are increasingly becoming available. Glueball masses of
$m(0^{++})=1611(30)(160)$ MeV;  $m(2^{++})=2232(220)(220)$ MeV and
$m(0^{-+})=2611(370)(260)$ MeV are  predicted where the second error is
an overall scale error. The quenched approximation  also gives
information on quark-antiquark scalar mesons and their mixing with
glueballs. This determination of the mixing in the quenched
approximation  is currently more powerful than attempting to determine
the mixed spectrum directly  in full QCD because the sea-quark masses 
realisable are rather heavy. There is also some lattice information  on
the hadronic decay  amplitudes of glueballs.

 For hybrid mesons, there will be no mixing with $q \bar{q}$ for 
spin-exotic states and these are the most useful predictions. The
$J^{PC}=1^{-+}$ state is expected at 10.81(25) GeV for $b$ quarks; 
4.19(15) GeV for $c$ quarks, 2.0(2) GeV for $s$ quarks and 1.9(2) GeV 
for $u,\ d$ quarks. Mixing of spin-exotic hybrids with
$q\bar{q}q\bar{q}$ or equivalently with meson-meson  is  allowed and
will modify the  predictions from the quenched approximation.

 Much activity is currently underway to explore the effects of sea
quarks  of ever decreasing mass. Future teraflops computing facilities
will  be essential to obtain quantitative results  for hadronic
spectroscopy.

\end{document}